\magnification=\magstep1
\baselineskip=13pt
\centerline{\bf QUANTUM DARWINISM AND ENVARIANCE\footnote*{\rm To appear in {\it Science and Ultimate Reality: From Quantum 
to Cosmos}, J. D. Barrow, P. C. W. Davies, and C. H. Harper, eds. 
(Cambridge University Press, 2004)}}
\bigskip
\centerline{Wojciech H. Zurek}
\medskip
\centerline{Theory Division, Mail Stop B210, LANL}
\centerline{Los Alamos, New Mexico 87545}
\bigskip
\centerline {\bf Abstract}
\medskip
\noindent Effective classicality of a property of a quantum system can be 
defined using redundancy of its record in the environment. This allowes
quantum physics to approximate the situation encountered in the classical world:
The information about a classical system can exist independently from its state.
In quantum theory this is no longer possible: In an isolated quantum system
the state and the information about it are inextricably linked, and any 
measurement may -- and usually will -- reset that state. However, when 
the information about the state of a quantum system is spread throughout 
the environment, it can be treated (almost) as in classical physics -- as 
(in effect) independent from the state of the open quantum system of interest. 
This is a central idea that motivates the {\it quantum Darwinism} approach 
to the interpretation problem. Quantum Darwinism differs from the traditional 
approach suggested by the von Neumann model of quantum measurement and offers 
a new perspective on the emergence of the everyday classical reality that 
is complementary to the one suggested by decoherence: Selection of preferred 
states occurs as a result of the `selective advertising', a proliferation of 
the information about the stable pointer states throughout the Universe. 
This view of the emergence of the classical can be regarded as (a Darwinian) 
{\it natural selection} of the preferred states. Thus, (evolutionary) 
{\it fitness} of the state is defined both by its ability to survive intact 
in spite of the immersion in the environment (i.e., environment-induced
superselection is still important) but also by its propensity to create 
offspring -- copies of the information describing the state of the system 
in that environment. I show that this ability to `survive and procreate' 
is central to effective classicality of quantum states. Environment retains
its decohering role, but it also becomes a ``communication channel'' through 
which the state of the system is found out by the observers. In this sense, 
indirect acquisition of the information about the system from its environment 
allows quantum theory to come close to what happens in the classical physics: 
The information about a classical system can be ``dissociated'' from its state. 
(In the case of an isolated quantum system this is impossible -- what is known 
about it is inseparably tied to the state it is in.) I review key ideas of 
{\it quantum Darwinism} and investigate its connections with the 
{\it environment -- assisted invariance} or {\it envariance}, a recently 
identified symmetry exhibited by pairs of entangled quantum systems that 
is responsible for the emergence of probability (allowing, in particular,
a completely quantum derivation of the Born's rule) within the wholly quantum 
Universe. 

\bigskip

\noindent{INTRODUCTION}

Quantum measurement problem is a technical euphemism for a much deeper 
and less well - defined question: How do we, `the observers', fit within the
physical Universe? This issue is especially apparent in quantum physics 
because, for the first time in the history of science a majority (but not all) 
physicists seriously entertain the possibility that the framework for 
the ultimate universal physical theory is provided by quantum mechanics.

Quantum physics relevant for this discussion is (contrary to the common 
prejudice) relatively simple. By this I mean that some of the key features 
of its predictions can be arrived at on the basis of the overarching principles
of quantum theory and without reference to the minutae of other specific 
ingredients (such as the details of the forces). 
Quantum superposition principle is such an overarching principle of quantum
theory. It leads to predictions that seem difficult to reconcile with our 
perception of the familiar classical Universe of everyday experience. The aim 
of this paper is to show that the emergence of the classical reality can be 
viewed as a result of the emergence of the preferred states from within 
the quantum substrate thorough the Darwinian paradigm, once the 
{\it survival of the fittest quantum states and selective proliferation of 
the information about them} are properly taken into account. 

Measurement problem has been the focus of discussions of the interpretation
of quantum theory since its inception in its present form in the mid-1920's.
Two new ideas that are the focus of this paper -- quantum Darwinism$^{1-3}$ 
and envariance$^{1,4}$ -- were introduced very recently. Exploration
of their consequences has only started. This presentation provides a somewhat 
premature (and, consequently, rather speculative) ``preview" of
their implications. We shall start with the von Neumann model$^5$
of quantum measurements. It has provided the standard setting for the
exploration of the role of observers and information transfer since it was 
introduced in 1932. We shall then go 
on and describe how von Neumann's model is modified by the introduction
of the environment in the more modern treatements, and briefly review
consequences of decoherence and of the {\it environment - induced 
superselection} or {\it einselection} that settle some of the issues.

Quantum Darwinism and envariance rely on the presence of the 
environment. They explore a similar set of questions as the 
theory of decoherence and einselection$^{1,6-9}$, but from a very different
vantage point: Rather than limit attention to the consequences of the
immersion of the system ${\cal S}$ or of the apparatus ${\cal A}$ in 
the environment ${\cal E}$ on the state of ${\cal SA}$, the focus shifts 
to the effect of the state of ${\cal SA}$ (or more precisely, the to - be - 
classical observables of that object, including in particular the apparatus 
pointer ${\cal A}$) on the state of the environment. 

The study of decoherence already calls for a modification of von Neumann's 
model -- for the addition of the environment. Quantum Darwinism is another 
radical change -- a change of focus, of the subject of discourse.  It is based 
on the realisation that almost without exception we -- the observers -- 
acquire information about  ``measured systems'' or the state of the
``appratus pointers'' indirectly -- by monitoring the environment: It
correlates with the system as a result of decoherence, which is caused 
by the environment (in effect) monitoring ${\cal A}$ and / or ${\cal S}$. 

Monitoring by the environment is responsible for the negative selection, for 
destabilisation of the vast majority of the states in the relevant Hilbert 
spaces of the open systems. What is left are the preferred pointer states. 
This, in essence, is the environment- induced superselection {\it einselection}.
Quantum Darwinism is based on the observation that intercepting such 
`second hand' information about the system by measuring fragments of 
the environment makes only some of the states of the system of interest 
accessible. These states happen to be the preferred {\it pointer states} of 
${\cal S}$. The reason for their selection that is also `Darwinian': 
Pointer states are not only best in {\it surviving} the hostile environment, 
but are also best in {\it proliferating} -- throughout the rest of the Universe,
using environment as the medium -- the information about themselves. This 
allows many observers to simultaneously find out about the pointer states, 
and to do so indirectly, and therefore, without perturbing them any further
than decohrerence already did. Objective existence of pointer states of quantum 
systems can be accounted for in this way$^{1-3}$. Hence, quantum analogues 
of the Darwinian criterion of `fitness' can be seen in the (ein-)selection of 
`the classical'.

Envariance focuses on the origins of ignorance (and, hence, information) in
the quantum Universe. It leads to the definition of probabilities -- to the
completely quantum derivation$^{1,4}$ of the Born's rule. Again, introduction 
of the environment is essential in this argument. In its presence one can 
delineate what aspects of the state of a system (that is correlated with 
the environment) cannot be known to the observer. In this way -- by starting
from a quantum definition of ignorance -- the operational definition of 
probabilities can be obtained as a consequence of a quintesentially quantum 
sort of a correlation -- quantum entanglement. It is interesting to note that 
analogous derivation cannot be repeated classically. This is because in 
classical physics information about the state can be ``dissociated" from that 
state, while in quantum physics what is known about the state cannot be 
treated separately from the state. Consequently, in quantum physics it is
possible to know precisely the joint state of two systems, but be provably
ignorant about the states of the component subsystems.

Both of these themes -- quantum natural selection and envariance -- have 
benefited from the inspiration and support of John Archibald Wheeler. 
To begin with, one of the two portraits displayed prominently in John's office 
in Austin, Texas, was of Charles Darwin (the other one was of Abraham Lincoln). 
This was symptomatic of the role theme of evolution played in John's thinking 
about physics (see, e.g., Wheeler's ideas on the evolutionary origin of 
physical laws$^{10}$). While I was always fond of looking at the 
`natural world' in Darwinian terms, this tendency was very much encouraged 
by John's influence. It seems quite natural to look at the emergence 
of the classical as a consequence of a quantum analogue of natural selection. 
Last not least (and on a lighter note) while my wife Anna and I were
visiting John on his `High Island' summer estate in Maine, we were put up in
a cottage in which -- I was told -- James Watson wrote ``The Double Helix"...

While quantum Darwinism benefited from Wheeler's boldness and encouragement,
{\it envariance} bears a more direct Wheeler {\it imprimitur}: Late in the year
1981 John and I were putting finishing touches on {\it Quantum Theory and 
Measurement}$^{11}$, and that included writing a section on {\it Further 
Literature}. At that time I was fascinated with the idea that quantum states
of entangled systems are in a sense relative -- defined with respect to one
another. Thus, John has caught me speculating {\it ... Zurek notes that 
``Nothing can keep one from thinking about [the two spins in a singlet] as 
the measured system and ... a quantum apparatus. [In that language] ... 
spin-system always points in the direction which is opposite to the direction 
of the ... spin-apparatus. This is a definite, ``coordinate - independent'' 
statement.''}  John overcame my reluctance and included these musings about 
`the relativity of quantum observables' (see p. 772 of Ref. 11). These very 
same ideas have recently -- and after a long gestation period -- begun to 
mature into a new way of looking at information and ignorance in the quantum 
context. Derivation of Born's rule based on the symmetries anticipated in that 
twenty - years old passage is presented in this paper in the sections on 
envariance. I am sure that this result is just a ``tip of the iceberg'', 
and I am convinced that envariance will prove to be a useful way of looking 
at various quantum issues of both fundamental and practical significance.

\bigskip
\noindent QUANTUM MEASUREMENT: VON NEUMANN'S MODEL

The traditional statement of the measurement problem goes back to 
von Neumann$^5$, who has analysed unitary evolutions that take initial state 
$|\psi_{\cal S}\rangle$	of the system and $|A_0\rangle$ of the apparatus
into and entagled joint state $|\Psi_{\cal SA}\rangle$:
$$ |\psi_{\cal S}\rangle |A_0\rangle = (\sum_k a_k |s_k\rangle) |A_0\rangle
\ \longrightarrow  \ \sum_k a_k |s_k\rangle |A_k\rangle = |\Psi_{\cal SA}\rangle \ .
\eqno(1)$$
Von Neumann has realised that while $|\Psi_{\cal SA}\rangle$ exhibits the
desired correlation between ${\cal S}$ and ${\cal A}$, the unitary
pre-measurement (as the `conditional dynamics' step described by Eq. (1) 
is often called) does not provide a satisfactory account of a `real world'
measurements. There are two reasons why Eq. (1) falls short of that goal: 
They are respectively identified as ``basis ambiguity'' and ``collapse of 
the wavepacket''. 

Basis ambiguity$^{12}$ is a direct consequence of the superposition principle:
According to it, one can rewrite an entangled bipartite state such as 
$|\Psi_{\cal SA}\rangle$ of Eq. (1) in an arbitrary basis of one of the two 
subsystems (say, ${\cal S}$) and then identify the corresponding basis of 
the other (i.e., the apparatus ${\cal A}$). That is:
$$ |\Psi_{\cal SA}\rangle = \sum_k a_k |s_k\rangle |A_k\rangle =
\sum_k b_k |r_k\rangle |B_k\rangle = \dots \  \ ,  \eqno(2)$$
where $\{|s_k\rangle\}$ and $\{|r_k\rangle\}$ (as well as $\{|A_k\rangle\}$ and
$\{|B_k\rangle\}$) span the same Hilbert space ${\cal H_S}$ (${\cal H_A}$),
while $a_k$ (and $b_k$) are complex coefficients.

Basis ambiguity can be also regarded as a consequence of entanglement. It is 
troubling, as it seems to imply that not just the outcome of the measurement, 
but also the set of states that describe the apparatus is arbitrary. Hence, any 
conceivable superposition (including the counterintuitive ``Schr\"odinger cat'' 
states$^{13}$) should have an equal right to be a valid description of a real 
apparatus (or a real cat) in a completely quantum Universe. This is blatantly 
at odds with our experience of the macroscopic objects (including, for instance,
states of the pointers of measuring devices) which explore only a very limited 
subset of the Hilbert space of the system restricted 
to the familiar, localized, effectively classical states.

The problem with the ``collapse of the wavepacket'' would persist even if one 
were to somehow identify the preferred basis in the Hilbert space of the 
apparatus, so that prior to observer's contact with ${\cal A}$ one could 
be at least certain of the `menu' of the possible outcome states of 
the apparatus, and the basis ambiguity would disappear. For, in the
end, we perceive only one of the possibilities, the actual outcome of the
measurement. ``Collapse'' is the (apparently random) selection of just one of 
the positions on the `menu' of the potential outcomes with the probability given
by Born's rule$^{14}$. 

Von Neumann discussed two processes that address the two aspects of the 
``quantum measurement process'' described above. While his investigation
preceded the famos EPR paper$^{15}$, and, hence, appreciation of the role of
entanglement (which is behind the basis ambiguity),  he has nevertheless
postulated {\it ad hoc} a non-unitary `reduction' from a pure state into 
a mixture:
$$ |\Psi_{\cal SA}\rangle \langle\Psi_{\cal SA}| \longrightarrow  
\sum_k |a_k|^2 |s_k\rangle \langle s_k| |A_k\rangle \langle A_k| 
= \rho_{\cal SA} \ . \eqno(3)$$
This process would have (obviously) selected the preferred basis. 
Moreover, von Neumann has also speculated about the nature of the next step 
-- the collapse, i.e., the perception, by the observer, of a unique 
outcome. This could be represented by a non-unitary transition, e.g. in 
a particular run of the experiment when state $\{|s_{17}\rangle\}$ is found:
$$\sum_k |a_k|^2 |s_k\rangle \langle s_k| |A_k\rangle \langle A_k| 
\longrightarrow |s_{17}\rangle\langle s_{17}| |A_{17}\rangle \langle A_{17}| \ .
\eqno(4)$$
In the collapse, the probability of any given outcome is given by Born's rule, 
$p_k = |a_k|^2$. Von Neumann has even considered the possibility that collapse 
may be precipitated by the conscious observers. This `anthropic' theme was 
later taken up by the others, including London and Bauer$^{16}$ and Wigner$^{17}$.

The aim of this paper is to investigate -- and where possible to settle --
open questions within the unitary quantum theory {\it per se}, without
invoking any non-unitary or anthropic {\it deus ex machina} such as Eqs. (3) or
(4) above.

\bigskip

\noindent DECOHERENCE AND EINSELECTION

Contemporary view (dubbed even ``the new orthodoxy''$^{18}$) is that 
the solution of the measurement problem -- and, in particular, the resolution
of the issues described above in the context of von Neumann's original
model -- requires a more realistic account of what actually happens during
a measurement: While von Neumann has treated the ${\cal SA}$ pair as isolated 
from the rest of the Universe, the discussions over the past two decades
have paid a lot of attention to the consequences of the immersion of 
the apparatus (and, more generally, of all the macroscopic objects) in their
environments$^{1-4,6-9,12,19-21}$. 

When the impossibility of perfect isolation of ${\cal A}$ is recognised, the
solution of the basis ambiguity problem can be obtained$^{1-3,12,19-21}$. 
Preferred basis -- candidate for the classical
basis in the Hilbert space of the system coupled to the environment --
is induced by the interaction with ${\cal E}$, which can be regarded as
monitoring by the environment of ${\cal A}$. The resulting transfer of
information is selective. Thus, observer who is in turn monitoring the 
environment to find out about the state of ${\cal A}$ will obtain only 
censored information: He will be able to readily find out about the preferred
pointer states of the system, but but it will be next to impossible to
find out about their superpositions. Egalitarian principle of superposition 
-- the cornerstone of quantum mechanics -- is grossly violated in such 
``open'' quantum systems. Different quantum states exhibit very different 
degree of resilience in presence of the interaction with the outside. Thus, 
the question about effective classicality is answered by the study of the
relative stability. {\it States that exist are the states that persist} 
is one of the tenets of the {\it existential interpretation}$^{8,9}$: 

Preferred pointer states are -- in contrast to arbitrary superpositions,
which, in accord with the superposition principle, have equal right to 
inhabit Hilbert space of an {\it isolated} system -- resilient to 
the entangling interaction with the environment. Hence, they maintain 
their identity -- their ability to faithfully represent the system. 
Selection of the preferred set of resilient pointer states in the Hilbert space
is the essence of the environment - induced superselection (einselection).
It is caused by a (pre-)measurement - like unitary evolution in which the
environment ${\cal E}$ becomes entangled with the apparatus:
$$ |\Psi_{\cal SA}\rangle |e_0\rangle = (\sum_k a_k |s_k\rangle |A_k\rangle)
|e_0\rangle \ \longrightarrow \ \sum_k a_k |s_k\rangle |A_k\rangle |e_k\rangle
= |\Phi_{\cal SAE}\rangle \ . \eqno(5)$$
When the state of the environment contains an accurate record of the outcome,
so that $|\langle e_k | e_l \rangle|^2 = \delta_{kl}$, the density matrix 
of the apparatus -- system pair acquires the desired form, as can be seen by
tracing out the environment:
$$ \rho_{\cal SA} = {\rm Tr}_{\cal E} |\Phi_{\cal SAE}\rangle 
\langle \Phi_{\cal SAE}| =
\sum_k |a_k|^2 |s_k\rangle \langle s_k| |A_k\rangle \langle A_k| .
\eqno(6)$$
This is clearly what is needed to solve the basis ambiguity problem (compare 
with Eq. (3) above). Moreover, it has been by now confirmed in model 
calculations and corroborated by experiments that the preferred pointer basis 
will habitually appear on the diagonal of the density matrix describing 
${\cal A}$ after a decoherence time (which is very short for the macroscopic
systems). The question, however, can be raised about the justification of 
the trace operation: The form of the density matrix relies on Born's rule$^{5}$.
Moreover, Eq. (6) gets only half of the job done: Eq. (3) -- the collapse 
-- still needs to be understood.

Within the context of decoherence and einselection both of these questions
-- basis ambiguity and collapse -- can be (albeit to a different degree) 
addressed. It is by now largely accepted (as a result of extensive studies 
of specific models) that under a reasonable set of realistic assumptions 
preferred basis of an apparatus pointer (or of selected observables of 
any open macroscopic system) does indeed emerge. Thus, quantum 
entanglement (present after the pre-measurement, Eq. (1)) will give way
to a classical correlation between ${\cal S}$ and ${\cal A}$, with the 
same preferred pointer basis $\{|A_k\rangle\}$ habitually appearing on the
diagonal of $\rho_{\cal SA}$. This takes care of the basis ambiguity.

This conclusion, however, crucially depends on the trace operation, which is
justified by employing Born's rule -- an important part of the quantum 
foundations, that is often regarded as an independent axiom of quantum
theory intimately tied with the process of measurement. One may (as many
have) simply accept Born's rule as one of the axioms. But it would be clearly 
much more satisfying to derive it. This will be our aim in the discussion of 
envariance.

The other outstanding issue is the apparent collapse and -- in particular -- the
{\it objectivity} of effectively classical (but, presumably, ultimately quantum)
states. That is, classical states can be simply `found out' by an observer
who is initially completely ignorant. This is not the case for quantum states:
Ideal measurement always yields an eigenvalue of the measured observable.  
Hence, it selects its (possibly degenerate) eigenstate. When the system 
does not happen to be in one of the eigenstates of the observable selected 
by the observer its measurement will perturb the state of the system
by resetting it to one of the eigenstates of what is being measured. Yet, 
in our everyday experience we never have to face this problem: Somehow, at 
the macroscopic level classical reality is a fact of life: We find out about 
the rest of the Universe at will, without having to worry about what does (and 
what does not) {\it exist}. We start by addressing this second issue of the
emergence of {\it objectivity}. 

\bigskip
\noindent QUANTUM DARWINISM

A part of the paradigm of ``quantum measurements'' that is shared not just by 
von Neumann's model, but by most of the other approaches to the interpretation
of quantum mechanics is the belief that we -- the observers -- acquire 
information about quantum systems directly, i.e., by interacting with them.
As was pointed out some time ago$^{8,9}$, this is never the case. 
For instance, a vast majority of our information is acquired visually. 
The information we obtain in this way does not concern photons, although
our eyes act as photon detectors: Rather, photons play the role of
carriers of information
about objects that emitt or scatter them. Moreover, we obtain all the 
information by intercepting only a small fraction of photons emitted by or
scattered from the object of interest with our eyes. Thus, many more copies 
of the same information must be carried away by such photon environment. 
Upon reflection one is led to conclude that essentially the same scheme 
(but involing different carriers of information) is the rule rather than 
exception. Measurements carried out on the macroscopic objects are invariably 
indirect, and carriers of information always ``fan out'' most of the copies 
of the `data', spreading it throughout the Universe. Observers use a fraction 
of the same environment that causes decoherence as a channel, to obtain 
information about the system of interest.

This distinction between direct and indirect acquisition of information
may seem inconsequential. After all, replacing a direct measurement with 
an indirect one only extends the `von Neumann chain'$^5$.  The overall state 
has the form of Eq. (5) and is still pure, with all of the potential outcomes 
present, superficially with no evidence of either Eq. (3) or the ``collapse''
of Eq. (4). Still, we shall show that when this situation is
analysed from the point of view of the observer, most (and perhaps all) of 
the symptoms of classicality emerge. 

To investigate a simple model of this situation we consider obvious 
generalisation of Eq. (5) we have used to describe decoherence:
$$ |\Psi_{\cal SA}\rangle \otimes_{n=1}^{\cal N} |e_0^{(n)}\rangle = 
(\sum_k a_k |s_k\rangle |A_k\rangle) \otimes_{n=1}^{\cal N}|e_0^{(n)}\rangle 
 \longrightarrow   \sum_k a_k |s_k\rangle |A_k\rangle \otimes_{n=1}^{\cal N}
|e_k^{(n)}\rangle = |\Phi_{\cal SAE^{N}}\rangle. \eqno(7)$$
There are ${\cal N}$ environment subsystems here. The assumption is that they
exist, and that they can be (like photons) accessed one at a time. 

We first note that enlarging this composite environment ${\cal E^{\cal N}}$ of 
Eq. (7) is absolutely irrelevant from the point of view of its effect on
the density matrix of the `object of interest', $\rho_{\cal SA}$.
For, when either a simple environment of Eq. (5) or the multiple environment
of Eq. (7) are traced out, the same $\rho_{\cal SA}$ of Eq. (6) will obtain.
So what (if anything) have we gained by complicating the model? Whatever
it is, obviously cannot be inferred from the state of ${\cal SA}$ alone. Yet,
in classical physics the state of ``the object of interest'' was all that
mattered! So where should we look now?

The inability to appreciate the implications of the difference between these
two situations is indeed firmly rooted in the `classical prejudice' that the
information about the system is synonymous with its state, but that the  
presence of that information is physically irrelevant for that state. This 
classical belief in the analogue of the `separation of church from state' 
is untenable in the quantum setting. For starters, there is `no information 
without representation'$^{21}$! 

Guided by our previous considerations, we shift attention from the state
of the object of interest (the ${\cal SA}$ pair) to the record of its state
in the environment. Now there is our difference! Instead of a single (fragile) 
record of the state of the system we now have many identical copies. How many? 
The preferred states $\{|A_k\rangle\}$ of the apparatus have left ${\cal N}$ 
imprints on the environment. This is easily seen in the example above, and can 
be quantified by one of the versions of the {\it redundancy ratio}$^{1-3}$, 
which in effect count the number of copies of the information
about the object of interest spread throughout the environment.

One definition of the redundancy ratio starts with redundancy defined 
in terms of the familiar mutual information -- a measure of correlation 
between the fragment of the environment ${\cal E}^{(n)}$ and the object 
of interest$^{1,2}$. This leads to;
$$ I({\cal S:E}^{(n)}) = H({\cal S}) + H({\cal E}^{(n)}) - H({\cal S,E}^{(n)}) \ , \eqno(8)$$ 
where $H({\cal S})$, $H({\cal E}^{(n)})$, and $H({\cal S,E}^{(n)})$ are the
relevant individual and joint entropies. Above, we have also replaced 
${\cal SA}$ of Eq. (5) by a single object to simplify notation, 
and to emphasize that this approach applies in general -- and not 
just in measurement situations. Various entropies can be defined in several 
ways using obvious reduced density matrices of the relevant subsystems of 
the whole$^{1,2,22}$. Redundancy can be then estimated as:
$$ {\cal I^{(N)}} = \sum_{k=1}^{\cal N} I({\cal SA:E}^{(k)}) \ . \eqno(9)$$
The physical significance of redundancy in the context of our discussion is 
similar to its import in the classical information theory$^{23}$: Redundancy 
protects information about the object of interest. From the point of view of 
the interpretation of quantum theory, this implies, for example, that many 
different observers can find out the state of the object of interest 
independently -- by measuring different fragments of the environment. 
This is how -- I believe -- states of the ultimately quantum but macroscopic 
objects in the world of our everyday experience acquire their {\it objective
existence}$^{1-3}$. 

However, viewed in a Darwinian fashion, redundancy ratio has also a different
significance: It provides, in effect, a measure of the number of ``offspring''
of the state in question. Thus, in the ideal case we have considered above
proliferation of information has led to ${\cal N}$ descendants of the original
state of the apparatus. The redundancy ratio in the example given above
is:
$$ {\cal R} = {\cal I^{(N)}}/H({\cal S}) = {\cal N} \ . \eqno(10)$$
Both the prerequisites for, and the consequences of high redundancy have 
significance that is best appreciated by invoking analogies with the 
``survival of the fittest". To begin with, a state that manages to spread 
many imprints of its `genetic information' throughout the environment must 
survive long enough -- must be resistant to the perturbations caused by 
the environment. This points immediately to the connection with the pointer 
states$^{12}$ -- they remain unperturbed by decoherence. But this is in 
a sense just a different view of selection of the preferred states, which
does not capitalise on the measure of their fecundity we have introduced
above.

Darwinian analogy recognizes that proliferation of certain information 
throughout the environment makes its further proliferation more likely. 
This is best seen in a still more realistic 
extension of the models of the environments we have considered so far: 
Suppose that in addition to the immediate environments ${\cal E}^{(k)}$
there are also distant environments $\varepsilon^{(l)}$, which do not
interact directly with ${\cal S}$ but interact with 
the immediate environments through interaction that is local -- i.e., that 
allows individual subsystems of the immediate environment to become correlated
with individual subsystems of the distant environment. Then it is easy
argue that the only information about ${\cal S}$ that can be passed
along from ${\cal E}$'s to $\varepsilon$'s will have to do with the preferred
pointer states: Only locally accessible information$^{22}$ can be passed
along by such local interactions. Indeed, this connection between the 
selection of the preferred basis and redundancy was noted already some time
ago$^{19,24}$.

We note in passing that there is an intimate relation between this 
necessity to make a selection of preferred states in the setting that involves
``fan-out'' of the information and the {\it no-cloning theorem}$^{25}$, which,
in effect, says that copying implies a selection of a preferred set of states
that are copied. 
We also note that all of the above considerations depend on the ability to 
split the Universe as a whole into subsystems. This -- as was already
noted in the past --- is a prerequisite of decoherence. Moreover, problems
of interpretation of quantum physics do not arise in a Universe that does not
consist of subsystems$^{1,8,9}$. 

\bigskip
\noindent ENVIRONMENT - ASSISTED INVARIANCE

{\it Envariance} is an abbreviation for {\it environment - assisted invariance},
the peculiarly quantum symmetry exhibited by the states of entangled
quantum systems. To explain it we consider a state vector describing system 
${\cal S}$ entangled (but no longer interacting) with the environment 
${\cal E}$. The joint state can be always written in the Schmidt basis:
$$ |\psi_{\cal SE}\rangle = \sum_k^N \alpha_k |s_k\rangle |\varepsilon_k\rangle \ . \eqno(11a)$$
For, even when the initial joint state is mixed, one can always 
imagine purifying it by enlarging the environment. As the environment
no longer interacts with the system, probabilities of various states of the
system cannot be -- on physical grounds -- influenced by such purification.
In writing Eq. (11) we assumed that such purification was either 
unnecessary or was already carried out.

Environment - assisted invariance refers to the fact that there is a family
of unitary quantum transformations $U_{\cal S}$ that act on a system alone, 
and are non-trivial, so that $U_{\cal S} |\psi_{\cal SE}\rangle \neq
|\psi_{\cal SE}\rangle$, but their effect can be undone by acting solely
on ${\cal E}$. Thus, for any $U_{\cal S}$ that has Schmidt states as 
eigenstates one can always find $U_{\cal E}$ such that:
$$ U_{\cal E} (U_{\cal S}  |\psi_{\cal SE}\rangle ) =
|\psi_{\cal SE}\rangle \eqno(12)$$
This is evident, as unitaries with Schmidt eigenstates acting on ${\cal S}$
will only rotate the phases of the coefficients $\psi_{\cal SE}$. But these 
phases can be also rotated by acting on ${\cal E}$ alone. Hence, 
transformations of this kind are envariant. It turns out that
envariant transformations always have Schmidt eigenstates$^1$.

In the spirit of decoherence we now focus on the system alone. Clearly, 
for an observer with no access to ${\cal E}$, system must be completely 
characterised by the set of pairs $\{|\alpha_k|,~|s_k\rangle\}$:  Only the 
absolute values of the coefficients can matter since phases of $\alpha_k$ 
can be altered by acting on ${\cal E}$ alone, and ${\cal E}$ is causally 
disconnected from ${\cal S}$. Thus, in the case when all $|\alpha_k|$ 
are equal; 
$$ |\bar \psi_{\cal SE}\rangle = \sum_k^N |\alpha| e^{- i \varphi_k} 
|s_k\rangle |\varepsilon_k\rangle \ , \eqno(11b)$$
any orthonormal basis is obviously Schmidt, and we can use envariance
to re-assign the coefficients to different states 
$\alpha_k \rightarrow \alpha_l$, $\alpha_l \rightarrow \alpha_k$, etc. 
Such swapping leaves the description of the system invariant: The coefficients
can differ only by the phase, and we have proved above that phases of the
Schmidt coefficients cannot influence probabilities of the system alone$^{1,4}$.
(Indeed, if this was possible, faster that light communication would be also
possible, as the reader can easily establish by extending the above argument.)

It is now evident that the probabilities of all $k$'s must be equal. Hence, 
assuming the obvious normalisation, they are given by:
$$ p_k=1/N \ . \eqno(13a)$$
Moreover, a collection of a subset of $n$ amongst $N$ 
mutually exclusive events (orthogonal states) has the probability:
$$ p_{k_1 \vee  k_2 \vee ... \vee \ k_n} = n / N  \ . \eqno(13b)$$
These results were easy to arrive at, but we have started with very strong
assumption about the coefficients. 

The case when $|\alpha_k|$ are {\it not} equal is of course of interest. 
We shall reduce it to the case of equal coefficients by extending the
Hilbert space of the environment. In the process we shall recover 
Born's rule $p_k = |\alpha_k|^2$. This will also provide a firmer foundation
for the decoherence approach which untill now uses Born's rule 
to justify its reliance on reduced density matrices. We note that we
have, in a sense, already gone half way in that direction: Phases in the Schmidt
decomposition have been already shown to be irrelevant, so the 
probabilities must depend on the absolute values of the coefficients.
We still do not know in what specific function is this dependence
embodied.

To illustrate the general strategy we start with an example involving 
a two-dimensional Hilbert space of the system spanned by states 
$\{|0\rangle,|2\rangle\}$ and (at least) a three-dimensional Hilbert space 
of the environment: 
$$ |\psi_{\cal SE} \rangle \ = \ (\sqrt 2 |0\rangle|+ \rangle \ \ + \ \  |2\rangle|2\rangle)/\sqrt 3\ . \eqno(15a)$$
The state of the system is on the left, and 
$|+\rangle=(|0\rangle+|1\rangle)/\sqrt 2$ exists 
in the (at least two-dimensional) subspace of ${\cal E}$ orthogonal to 
the environment state $|2\rangle$, so that 
$\langle0|1\rangle=\langle0|2\rangle=\langle1|2\rangle=\langle+|2\rangle=0$. 
To reduce this case to the case of Eq. (11b) we extend $|\psi_{\cal SE}\rangle$ 
above to a state 
$|\phi_{\cal SEE'}\rangle$ with equal coefficients by acting only on 
the causally disconnected ${\cal E}$ (which implies that probabilities 
we shall inferr for ${\cal S}$ could not have changed). This can be done 
by allowing a {\tt c-shift} act between ${\cal E}$ and ${\cal E'}$ so that
(in the obvious notation) $|k\rangle|0'\rangle \Rightarrow |k\rangle|k'\rangle$, and;
$$ |\psi_{\cal SE}\rangle|0\rangle = {{\sqrt 2 |0\rangle|+\rangle|0'\rangle + |2\rangle|2\rangle|0'\rangle} \over \sqrt 3}
\Longrightarrow 
\bigl(\sqrt 2 |0\rangle{{|0\rangle|0'\rangle + |1\rangle|1'\rangle} \over \sqrt 2}  + |2\rangle|2\rangle|0'\rangle\bigr)/\sqrt 3
 \eqno(16a)$$
The cancellation of $\sqrt 2$ leads to:
$$ |\phi_{\cal SEE'}\rangle = (|0\rangle|0\rangle|0'\rangle + |0\rangle|1\rangle|1'\rangle 
 + |2\rangle|2\rangle|2'\rangle)/\sqrt 3 
\eqno(17a)$$
The phases are again irrelevant as they can be altered by manipulating 
${\cal E'}$ alone. Clearly, for the bipartite combination of ${\cal S}$ and 
${\cal E}$ the three orthonormal product states have coefficients with same 
absolute value and can be swapped. Hence, all of them must have the same 
probability. Thus, by Eq. (13a), probabilities of 
$|0\rangle|0\rangle, \ |0\rangle|1\rangle,$ and $|2\rangle|2\rangle$ 
are all equal to 1/3. Moreover, two of them involve state $|0\rangle$ 
of the system.  So, by Eq. (13b), the probability of $|0\rangle$ state of 
the system is twice the probability of $|2\rangle$. Consequently:
$$ p_0 = 2/3; \ \ p_2=1/3 \ . \eqno(18a)$$
Hence, in this special case -- but using ideas that are generally applicable --
I have derived Born's rule, i.e., demonstrated that entanglement leads to
envariance and this implies $p_k=|\alpha_k|^2$. It is straightforward (if a bit
notationally cumbersome) to generalise this derivation, and we shall do 
so in a moment. But the basic idea is already apparent and worth 
contemplating before we proceed with a general case (where the main
point is somewhat obscured by notation). 

\bigskip
\noindent ENVARIANCE, IGNORANCE, AND INFORMATION

The above derivation of probabilities in quantum physics is very much 
in the spirit of the ``ignorance interpretation'', but in the quantum 
context it can be carried out with an important advantage: In the classical 
case observers assume that 
an unknown state they are about to discover exists objectively prior
to the measurement, and that the ignorance allowing for various swappings
reflects their ``subjective lack of knowledge''. Indeed, the clash between this subjectivity of information on one hand and its obvious physical significance
on the other has been a source of a long - standing friction distilled
into the Maxwell's demon paradox. {\it In quantum 
theory ignorance can be demonstrated in an objective fashion, as
a consequence of envariance of a state perfectly known as a whole.} Above,
${\cal SEE'}$ is pure. Quantum complementarity enforces ignorance of the
states of the parts as the price that must be paid for the perfect knowledge 
of the state of the whole.

It seems ironic that a natural (and a very powerful) strategy to 
justify probabilities rests -- in quantum physics -- on a more objective 
and secure foundation of perfectly known {\it entangled} pure states 
than in the deterministic classical physics: When the state of the observers 
memory $|\mu\rangle$ is not correlated with the system, 
$$|\Psi_{\mu {\cal SE}}\rangle \sim |\mu\rangle \sum_k |s_k\rangle |\epsilon_k\rangle \eqno(20)$$
and the absolute values of the coefficients in the Schmidt decomposition 
of the entangled state describing ${\cal SE}$ are all equal, and ${\cal E}$
cannot be accessed, the resulting state of ${\cal S}$ is {\it objectively invariant}
under all {\it local} measure - preserving transformations. Thus, with 
no need for further excuses, probabilities of 
events $\{|s_k\rangle\}$ must be -- prior to measurement -- equal.

By contrast, after observer (pre)measures the system, the overall state;
$$|\Phi_{\mu {\cal SE}}\rangle \sim \sum_k |\mu_k\rangle |s_k\rangle |\epsilon_k\rangle \eqno(21)$$
obtains, with the correlation between his record $|\mu_k\rangle$ and the system 
state $|s_k\rangle$ allowing him to inferr the state of the system from his
record state. The invariance we have appealed 
to before is substantially restricted: Correlated pairs $ |\mu_k\rangle |s_k\rangle $
can be no longer separated and have to be permuted together. Thus, to a 
friend of the observer, all outcomes remain equiprobable, but to the `owner
of the memory $\mu$' his state is in part described by what he has found out
about the system. Consequently, $|\mu_k\rangle$ implies $|s_k\rangle$ and 
the probability conditioned on observers own state in the wake of the perfect
measurement is simply $p_{ s_l | \mu_k} =\delta_{lk} $. 
Conditional probability in quantum theory emerges as an objective 
consequence of the relationship between the state of the observer 
and the rest of the Universe, as the combined state under consideration 
(and not just the ill-defined and dangerously subjective ``state of observers 
knowledge'' about a ``definite but unknown classical state'') is invariant 
in a manner that allows one to deduce equality of probabilities much more 
rigorously, directly and without the copious apologies required 
in the classical setting.

We note that the above discussion of the acquisition of information owes 
a great deal to Everett$^{26}$. The collapse occurrs on the way from Eq. (20) 
to Eq. (21). Envariance has given us a new insight into the nature of collapse: 
It is the extent of the correlations -- the proliferation of information --
that is essential in determining what states of quantum systems can  
be perceived by observers. When an envariant swap can be carried out on 
the ${\cal SE}$ pair, without involving the state of the observer 
(see Eq. (20)), he is obviously ignorant of the state of ${\cal S}$. By 
contrast, a swap in Eq. (21) would have to involve the state of the observer. 
This is because the information he has acquired is inscribed in the state of 
his own memory. (There is {\it no information without representation}$^{21}$.)
In this sense, envariance extends the {\it existential interpretation}$^{8,9}$
introduced some time ago to deal with the issue of collapse.

\medskip

\noindent BORN'S RULE FROM ENVARIANCE -- GENERAL CASE

To discuss the general case we start with the state:
$$ |\Psi_{\cal SE}\rangle = \sum_{k=1}^N \sqrt{m_k \over M} |s_k\rangle|\epsilon_k\rangle \ ,
\eqno(15b)$$
where $M = \sum_{k=1}^N m_k$ assures normalisation. As the coefficients are
commensurate, and as we assume that the Hilbert subspaces of ${\cal E}$
corresponding to different $k$ are at least $m_k$ dimensional,
appropriate {\tt c-shift}$^{1,2}$:
$$ |\epsilon_k\rangle|\varepsilon '\rangle =  
({1 \over \sqrt{m_k}  } \sum_{l_k=1}^{m_k}|\varepsilon_{l_k}\rangle)
~|\varepsilon '\rangle \   \Longrightarrow 
\ {1 \over \sqrt {m_k}} \sum_{l_k=1}^{m_k} |\varepsilon_{l_k}\rangle
|\varepsilon_{l_k} '\rangle $$
that couples ${\cal E}$ with at least as large ${\cal E'}$ yields:
$$ |\Psi_{\cal SE}\rangle|\varepsilon\rangle \Longrightarrow M^{-1} \sum_{k=1}^N |s_k\rangle
(\sum_{l_k}^{m_k}  |\varepsilon_{l_k}\rangle  |\varepsilon_{l_k} '\rangle) \eqno(16b)$$
Here, in contrast to (16a), we have immediately carried out 
the obvious cancellation,
($\sqrt{m_k} |\epsilon\rangle = \sum_{l_k}^{m_k} |\varepsilon_{l_k}\rangle $). 
It follows as a direct consequence of the relation between the states 
$|\epsilon_k\rangle$ and their Fourier-Hadamard transforms 
$|\varepsilon_k\rangle$.

The resulting state can be rewritten in a simpler and more obviously 
invariant form:
$$ |\Phi_{\cal SEE'}\rangle = M^{-1} \sum_{j=1}^M |s_{k(j)}\rangle |\varepsilon_{j}\rangle  |\varepsilon_{j} '\rangle \eqno(17b)$$
where the environmental states are orthonormal, and the system state
is the same within different $m_k$ - sized blocks (so that the same 
state $|s_{k(j)}\rangle$ appears for $m_k$ different values of $j$,
and $\sum_{k=1}^N m_k=M$).

As before (see Eq. (17a)) phases are irrelevant because of envariance. Hence,
terms corresponding to different values of $j$ can be swapped, and -- by
Eq. (13) -- their probabilities are all equal to $1/M$. It follows
that:
$$ p_k = p({|s_k\rangle}) = m_k / M = |\alpha_k|^2 \eqno(18b)$$
in obvious notation. This, as promised, is Born's rule. When $|\alpha_k|^2$
are not commensurate, one can easily produce sequences of states that set up
convergent bounds on $p_{s_k}$ so that -- when the probabilities are assumed
to be continuous in the amplitudes -- the interval containing $p_k$ shrinks
in proportion to $1/M$ for large $M$.

We emphasize again that one could not carry out the basic step of our argument 
-- the proof of the independence of the probabilities from the  phases of 
the Schmidt expansion coefficients, Eq. (12) and below  -- for an equal
amplitude pure state of a single, isolated system. The problem with:
$$ |\psi\rangle \ = \ N^{-{1 \over 2}} \sum_k^N \exp( i \phi_k) |k\rangle $$
is the accessibility of the phases. Consider, for instance;
$ |\psi\rangle \ = \ (|0\rangle + |1\rangle - |2\rangle)/\sqrt3 $ and
$ |\psi'\rangle \ = \ (|2\rangle + |1\rangle - |0\rangle)/\sqrt3 $.
{\it In the absence of entanglement there is no envariance} and swapping 
of states corresponding to various $k$'s is detectable: Interference 
measurements (i.e., measurements of the observables with (phase-dependent)
Hadamard eigenstates $|1\rangle+|2\rangle; \ |1\rangle-|2\rangle,$ etc.) 
would have revealed the difference between $|\psi\rangle$ and $|\psi'\rangle$. 
Indeed, given an ensemble of identical {\it pure} states a skilled observer
should be able to confirm that they are pure and find out what they are.
Loss of phase coherence is needed to allow for the shuffling of the 
states and coefficients.

Note that in our derivation environment and einselection play an additional,
more subtle role: Once a measurement has taken place -- i.e., a correlation 
with the apparatus or with the memory of the observer was established 
(e.g., Eqs. (21) and (22)) -- one would hope that records will
retain validity over a long time, well beyond the decoherence timescale. 
Thus, a ``collapse'' from a multitude of possibilities to a single reality 
(implied by Eq. (22) above) can be confirmed by subsequent
measurements only in the einselected pointer basis.

With this in mind, it is easy to see that -- especially on the macroscopic
level -- the einselected states are the only sensible choice as outcomes: 
Other sets of states lose correlation with the apparatus (or with the memory 
of the observer) far too rapidly -- on the decoherence timescale -- 
to serve as candidate events in the sample space. 

We close this part of our discussion by calling reader's attention 
to the fact that the above derivation did not rely on -- or even invoke 
-- reduced density matrices, which are at the very foundation of 
the decoherence program. Indeed, we have used envariance to
{\it derive} Born's rule, and, hence, in a sense, to justify the form 
and the uses of the reduced density matrices. More extensive discussion of
this point shall be given elsewhere (Zurek, in preparation).

\medskip

\noindent SUMMARY AND CONCLUSIONS: QUANTUM FACTS

In spite of the preliminary nature of much of the above (which would 
seem to make ``Conclusions'' premature) we point out that if one were 
forced to attach a single label to the topics explored above, {\it quantum
facts} would be a possible choice. Quantum Darwinism approaches this theme 
directly: Quantum states, by their very nature, share epistemological and 
ontological role -- they are simultaneously a description of the state,
and `the dream stuff is made of'. One might say that they are {\it epiontic}. 
These two aspects may seem contradictory, but, at least in quantum
setting, there is a union of these two functions.

Quantum Darwinism puts forward a specific theory of how the ontic aspect
-- reliable classical existence states -- can emerge from the quantum
substrate. We shall not repeat the arguments already given in detail.
But one might sum up the key idea by pointing to the role of the redundancy:
Tenuous quantum facts acquire objective existence when the information
they about them is widely spread (and therefore becomes easily accessible). 
Approximate (exact) classicality obtains in the limit of a large (infinite)
redundancy. Redundancy is a measure of classicality.

Envariance is, by contrast, a way to capture the most tenuous aspect of the
quantum -- the ignorance (and, hence, the essence of what is epistemic: the
information). Quantum facts are the opposite of envariant properties.
{\it Quantum fact are invariant under envariance}.
Thus, in a sense, what we have accomplished is to ``corral'' the problem 
of the emergence of the classical from quantum states between two extremes:
The case -- exploited by quantum Darwinism -- where quantum facts become
solid and reliable, and the opposite, when some properties of these
states are envariant, and, therefore, demonstrably inconsequential.
Investigation, in terms of envariance and quantum Darwinism, of what lies 
inbetween these two extremes is still in its early stages.

Extensive comments on the manuscript by Harold Ollivier are greatly appreciated.

\bigskip
\bigskip
\noindent REFERENCES
\medskip
\item{1.} Zurek, W.~H., 2003, {\it Decoherence, Einselection, 
and the Quantum Origins of the Classical}, Rev. Mod. Phys. {\bf 75}, 715-765.

\item{2.} Zurek, W.~H., 2000, Ann. der Physik (Leipzig) {\bf 9}, 855.

\item{3.} Ollivier, H., Poulin, D., and Zurek, W.~H., 2003, Phys. Rev. Lett.,
submitted.

\item{4.} Zurek, W.~H., 2003, {\it Environment - Assisted Invariance, Ignorance,
and Probabilities in Quantum Physics}, Phys. Rev. Lett. {\bf 90}, 120404.

\item{5.} von~Neumann, J., 1932, {\it Mathematische Grundlagen der
Quantenmechanik}, (Springer, Berlin); reprinted 1981;
English translation by R.~T. Beyer, 1955: {\it Mathematical
Foundations of Quantum Mechanics}, (Princeton
University Press).

\item{6.} Giulini, D., E. Joos, C. Kiefer, J. Kupsch, I.-O.
Stamatescu, and H. D. Zeh, 1996, {\it Decoherence and the Appearance
of a Classica World in Quantum Theory} (Springer, Berlin).

\item{7.} Paz, J.-P., and W.~H. Zurek, 2001, pp. 533-614 in {\it Coherent
Atomic Matter Waves, Les Houches Lectures}, R. Kaiser, C. Westbrook, and 
F. David, eds. (Springer, Berlin).

\item{8.} Zurek, W.~H., 1993, Progr. Theor. Phys. {\bf 89}, 281--312.

\item{9.} Zurek, W.~H., 1998, Phil. Trans. Roy. Soc. Lond. {\bf A356}, 1793.

\item{10.} Wheeler, J.~A., 1983, in Wheeler and Zurek, Ref. 11.

\item{11.} Wheeler, J.~A., and W.~H. Zurek, 1983, {\it Quantum Theory
and Measurement}, (Princeton University Press).

\item{12.} Zurek, W.~H., 1981, Phys.\ Rev. {\bf D24}, 1516--1525.

\item{13.} Schr\"odinger, E., 1935a, Nautrwiss {\bf 23}, 807--812, 823--828, 
844--849 [English translation in Wheeler and Zurek, Ref. 11]; 1935b, 
Proc.\ Cambridge Phil.\ Soc. {\bf 31}, 555--563; 1936, Proc. Cambridge Phil. 
Soc. {bf 32}, 446-452.

\item{14.} Born, M., 1926, Zeits. Physik.

\item{15.} Einstein, A., B. Podolsky, and N. Rosen, 1935, Phys.\ Rev.
{\bf 47}, 777--780; [reprinted in Wheeler and Zurek, Ref. 11].

\item{16.} London, F., and Bauer, E., 1939, {\it La Th\'eorie de
l'Observation en M\'echanique Quantique} (Hermann, Paris);
English translation in Wheeler and Zurek, 1983.

\item{17.} Wigner, E.~P., 1963, Am. J. Phys. {\bf 31}, 6-15; reprinted
in Wheeler and Zurek, Ref. 11.

\item{18.} Bub, J., 1997, {\it Interpreting the Quantum World},
Cambridge University Press, Cambridge.

\item{19.} Zurek, W.~H., 1982, Phys.\ Rev. {\bf D26}, 1862--1880.

\item{20.} Zurek, W.~H., 1991, Physics Today\/ {\bf 44} (Oct.), 36--44.

\item{21.} Zurek, W.~H., 1994, Decoherence and the Existential Interpretation 
of Quantum Theory, or ``No Information Without Representation'',
pp. 341-350 of {\it From Statistical Physics to Statistical Inference and
Back}, P. Grassberger and J.-P. Nadal, eds. (Plenum, Dordrecht, 1994).

\item{22.} Ollivier, H., and W.~H. Zurek, 2002, Phys. Rev. Lett. {\bf 88},
017901; Zurek, W.~H., 2003, Quantum Discord and Maxwell's Demons,
Phys. Rev. A. {\bf 67}, 012320. 

\item{23.} Cover, T.~M, and J.~A. Thomas, 1991 {\it Elements of
Information Theory} (Wiley, New York).

\item{24.} Zurek, W.~H., 1983, in {\it Quantum Optics, Experimental
Gravitation, and Measurement Theory}, edited by P. Meystre and M. O.
Scully (Plenum, New York), p. 87.

\item{25.} Wootters, W.~K., and W.~H. Zurek, 1982, Nature, {\bf 299}, 802;
Dieks, D., 1982, Phys. Lett. {\bf 92A}, 271.

\item{26.} Everett, H., III., 1957a, Rev.\ Mod.\ Phys. {\bf 29}, 454--462; 
reprinted in Ref. 11; 1957b, {\it The theory of the universal
wave function},  Princeton University Dissertation; reprinted in
DeWitt, B.~S., and Graham, N., 1973, {\it The Many-Worlds
Interpretation of Quantum Mechanics} (Princeton University Press).

\end